\documentclass[12pt,thmsa]{article}
\usepackage{amssymb}

\input{tcilatex}
\begin{document}

\author{\textbf{Mohamed Tarek Hussein, Nabila Mohamed Hassan} \and \textbf{And
Naglaa Elharbi} \and \textit{Physics Department, Faculty of Education for
girls, Boghdadia,} \and \textit{Jeddah, KSA.} \textit{P.O.Box 9470 Jeddah
21413} \and \textbf{\ }}
\title{Time Evolution of Fast Particles During the Decay of Hadronic Systems }
\date{The Date }
\maketitle

\begin{abstract}
A phenomenological model is presented based on the formation of nuclear
thermodynamic system during the collision of heavy ions in the regime of
intermediate and high energy regions. The formulation and the dynamic
picture are determined by solving the Vlasov equation. The solution is
dressed in the form of a power series. The first term of which is the
equilibrium distribution in phase space. The rest, are time dependent
perturbation terms due to the multiple strong interactions inside the
system. The temperature gradient and the derivatives of the phase function
are calculated. The time dependence of the angular emission of the produced
particles is studied. It is found that particles emitted in the forward
direction are produced in the early stage of the reaction, far from the
equilibrium. Backward production comes in a later stage when the system
constituents undergo multiple cascade collisions.
\end{abstract}

\section{Introduction}

The particle production in heavy ion collisions was well represented by the
fireball model [1] at medium energy range, where the concept of global
equilibrium may be accepted. The fireball model was developed to fit
experimental data at higher energies. A local equilibrium was assumed in the
so called fire-streak model [2,3] that treats the variation across the
overlap region of the target and projectile in the amount of energy and
momentum that it deposited. The expression for calculating any observable
takes the form of a sum over a series of terms, each one of which concerns
to a local equilibrium and consists of a geometric, kinematics and
statistical factors. As the energy increases more, it is expected that
collision time becomes small enough so that particle emitted in the early
stage of the reaction possesses non-equilibrium characteristics. The density
function in phase space should be treated on the time scale to follow up the
time grow of the reaction. Many trials have been done in this concern. The
equation of motion can be reformulated to give it the appearance of
classical equation for the phase distribution function. In this
approximation, a local one body potential can be defined and the phase
distribution function may contain the same information as the one body
density matrix. This is the Hartree-Fock approximation [4]. The many body
physics enters only through the relation of the potential and the density.
One more approximation reduces the equation to completely a classical form
is to make a power series expansion of the one body potential and get the so
called Vlasov equation [5-7]. A situation that can be analyzed with the
Vlasov equation is the short time behaviour of the system subjected to an
impulsive force. If the potential is sufficiently weak, the solution of the
excited system may be treated by the quantum mechanical sum rules introduce
first by Fallieros [8] and Noble [9]. While it is not possible to integrate
the Vlasov equation in general, some insight may be given by expanding the
solution for small intervals of time. The starting point is the equilibrium
solution, which is perturbed by the impulsive potential. Another treatment
of the Vlasov equation depends on the theory of small oscillations in finite
system [10]. A closed expression has the appearance of Rayleigh`s variation
principle with a certain explicit form for the potential energy function.
The solution is represented in the form of a sum of an equilibrium function
plus a time dependent one which is assumed small compared with the first.
The motion is assumed to have a sinusoidal time dependence with frequency.
The variational principle was applied to estimate the frequencies of nuclear
vibrations of various multi-polarity. In this work a method is developed to
solve the Vlasov equation with reasonable approximations in a frame of a
time dependent thermodynamic model, which enables the calculation of light
and heavy particle spectra on the different reaction stages. The details of
the model are presented in section 2. Results and discussion are displayed
in section 3, and finally in section 4, we present conclusive remarks.

\section{The Model}

Let us consider the collision between a target nucleus T and a projectile
one P at a given impact parameter $\overline{b}$. The collision goes through
sequential stages. The first is a compression of the nuclear matter due to
the high energy interaction, forming a fireball with diffuseness surface on
the contrary of the fireball model assumptions [1] which support the concept
of the participant and spectator nucleons with pure cylindrical cut in the
nuclear matter. The nuclear matter is then treated as a heterogeneous
thermodynamic system. Multiple nuclear collisions run inside the fireball
which increases the energy density and allows the formation of quark gluon
plasma state[11-14]. This leads to the creation of new particles and
expansion of the system which gradually approaches the equilibrium state.
The last stage is the fireball decay. Particle emission from the fireball is
allowed at different points on the time scale of the reaction. Light created
particles are expected to be emitted on the early stage at narrow forward
cone angle, i.e. due to the first few collisions. The higher order
collisions draw the system towards the equilibrium state producing particles
in isotropic distribution in phase space. It is then convenient to consider
the state of equilibrium as a time reference of the reaction. Drawing back,
we may follow the historical grow of particle emissions on the time scale.
Hadronic matter inside the fireball is partially formed by the fast
projectile nucleons and the slow target ones. The relative projectile
density in this mixture is a very important parameter. It determines the
fireball parameters, the center of mass velocity, the temperature and the
temperature gradient inside the fireball matter. We use a Gaussian density
distribution [15] for nuclei of mass number $A<20$, while a Fermi density
for $A\succeq 20$. Consider a frame of reference coincides with the center
of the target nucleus in the Lab. system, then the relative projectile
density $\eta (r,b)$, at a given distance $r$ inside the fireball matter and
a given impact parameter b, is given by:

\begin{equation}
\eta (r,b)=\frac{\rho _{p}(r-b)}{\rho _{p}(r-b)+\rho _{T}(r-b)}\text{ \ \ }
\end{equation}

Where, for $A<20$

\begin{eqnarray}
\rho _{i}(\overline{r}) &=&A_{i}(\pi r)^{-3/2}\exp (-r^{2}/R_{i}^{2}) 
\nonumber \\
R_{i}^{2} &=&(3/2)\frac{<r^{2}>}{(1-1/A_{T})}\text{ \ \ \hspace{1in}}i=p,T
\end{eqnarray}

and for $A\succeq 20$

\begin{eqnarray}
\rho _{i}(\overline{r}) &=&A_{i}\rho _{o}[1+\exp (\frac{r-c}{d})]^{-1} 
\nonumber \\
c_{i} &=&1.19A_{T}^{1/3}-1.61/\text{\ }A_{T}^{1/3}\text{ \ }fm\text{\hspace{%
1in}}i=p,T \\
d &=&0.54\text{ \quad }fm  \nonumber
\end{eqnarray}

The local temperature $T(\overline{r})$at a position vector $\overline{r}$
is the solution of the thermodynamic energy conservation Eq.,

\[
\zeta _{cm}=3T+m\frac{K_{1}(m/T)}{K_{2}(m/T)} 
\]

\begin{equation}
\lbrack m^{2}+2\eta (1-\eta )mt_{i}]^{1/2}=3T+m\frac{K_{1}(m/T)}{K_{2}(m/T)}
\end{equation}
where m is the rest mass of the constituent particle of the nuclear medium
under investigation, $K_{1},K_{2}$ are the McDonalds functions of first and
second order [16] and $t_{i}$ is the incident kinetic energy per nucleon.
Eq.(4) valid for each type of particles forming the fireball. The
temperature is very sensitive to the form of the nuclear density. The
momentum distribution of the fireball nucleons in the center of mass system
is given by:

\begin{equation}
\frac{d^{2}N}{p^{2}dpd\Omega }=\frac{N}{4\pi m^{3}}\frac{\exp (-E/T)}{%
2(T/m)^{2}K_{1}(m/T)+(T/m)K_{0}(m/T)}
\end{equation}
The equilibrium energy distribution in the lab system is given by: 
\begin{equation}
f_{o}(E,r)=pE^{\prime }\frac{d^{2}N}{p^{\prime 2}dp^{\prime }d\Omega }
\end{equation}
The prime letters are defined in the center of mass system and relativistic
transformed as:

\begin{equation}
E^{\prime }=\gamma _{cm}(E_{L}-\beta _{cm}P_{L}\cos \theta _{L})
\end{equation}
where the center of mass velocity $\beta _{cm}$ is given by, 
\begin{equation}
\beta _{cm}\frac{P_{L}}{E_{L}}=\frac{\eta [t_{i}(t_{i}+2m)]^{1/2}}{m+\eta
t_{i}}
\end{equation}
Since particles emission is allowed before approaching the equilibrium
state, then it is convenient to use the Vlasov equation [4] to deal with the
particle energy spectra at any time of the reaction. The Vlasov Eq. has the
form,

\begin{equation}
\frac{df}{dt}=\frac{\partial f}{\partial t}+\frac{\overline{P}}{m}\cdot
\triangledown _{r}f-\nabla _{r}U\cdot \nabla _{p}f
\end{equation}
Where $U(r)$ is a scalar potential acting among the particles. Eq.(9) may be
solved under some approximations. First, we shall consider a pre-equilibrium
state where the time derivative $\frac{df}{dt}$ may be approximated as $%
(f-f_{0})/t_{c}$ . Where $f_{o}$ is the equilibrium distribution. Since we
are dealing with a state near equilibrium, so it is convenient to consider
that the rate of change of the function f is approximately equal to that of $%
f_{o}$. So we replace $f$ by $f_{o}$ in the RHS of Eq.(9). Moreover, let us
consider the particles as almost free so that we neglect the potential $U$
in this stage of approximation. Eq.(9) then becomes, 
\begin{equation}
f_{1}=f_{0}+t_{c}\frac{\overline{P}}{m}\cdot \overline{\triangledown }%
_{r}f_{o}
\end{equation}

\[
\ =f_{0}+t_{c}\frac{P}{m}\cos \theta \frac{\partial f_{o}}{\partial r} 
\]
$f_{1}$ is the first order approximation of the particle spectrum, $t_c$ is
the time interval required by the system to approach the equilibrium state $%
f_{o} $ , and $\theta $ is the scattering angle, the angle between the
direction of particle emission $\overline{P}$ and the radial direction $%
\overline{r}$. A second order approximation is obtained by using $f_{1}$
instead of $f$ \thinspace in the RHS of Eq.(9), so that,$\ $%
\begin{equation}
f_{2}\ =f_{0}+t_{c}\frac{P}{m}\cos \theta \frac{\partial f_{o}}{\partial r}%
+(t_{c}\frac{P}{m}\cos \theta )^{2}\frac{\partial ^{2}f_{o}}{\partial r^{2}}
\end{equation}
By the same analogy we get the recursion relation for the $n^{th}$ order
approximation as;

\[
f_{n}=f_{0}+\sum_{i=1}^{n}(t_{c}\frac{P}{m}\cos \theta )^{i}\frac{\partial
^{i}f_{o}}{\partial r^{i}} 
\]
so that the third and fourth order approximations are;

\begin{equation}
f_{3}\ =f_{0}+t_{c}\frac{P}{m}\cos \theta \frac{\partial f_{o}}{\partial r}%
+(t_{c}\frac{P}{m}\cos \theta )^{2}\frac{\partial ^{2}f_{o}}{\partial r^{2}}%
+(t_{c}\frac{P}{m}\cos \theta )^{3}\frac{\partial ^{3}f_{o}}{\partial r^{3}}
\end{equation}

\begin{equation}
f_{4}\ =f_{0}+t_{c}\frac{P}{m}\cos \theta \frac{\partial f_{o}}{\partial r}%
+(t_{c}\frac{P}{m}\cos \theta )^{2}\frac{\partial ^{2}f_{o}}{\partial r^{2}}%
+(t_{c}\frac{P}{m}\cos \theta )^{3}\frac{\partial ^{3}f_{o}}{\partial r^{3}}%
+(t_{c}\frac{P}{m}\cos \theta )^{4}\frac{\partial ^{4}f_{o}}{\partial r^{4}}
\end{equation}

\section{Results and discussion}

The predictions of the pre equilibrium model are applied to the Ne-U
collisions at 400 and 2100 A MeV. Assuming a frame of reference coincide
with the center of the target, and that the projectile is located at a
position $\overline{r}$, with an impact parameter $\overline{b}$ as shown in
Fig.(1). The relative projectile density $\eta (r,b)$ is calculated
according to Eq.(1). In Fig.(2) we demonstrate $\overline{\eta }(r,b)$
averaged over the whole range of impact parameter. The function $\overline{%
\eta }(r,b)$ shows a peak value of a height ~ 0.5 at a distance ~$%
R_{p}+R_{T}=10.7fm$, where the projectile and the target have equal
densities. According to the model assumptions, the nuclei have no sharp
surface density but instead, a diffuseness surface which extends the range
of the nuclear matter to about twice the sum of the nuclear radii. On the
other hand the geometrical factor represented by the size of the nuclear
matter has heavy weight near the origin and falls exponentially with $%
\overline{r}$ toward the surface as may be described by the tail of the
Gaussian distribution. The effective range, where the nuclear matter forming
the nuclear thermodynamic system has appreciable value is estimated to about 
$1.5(R_{p}+R_{T})$. The parameter $\eta $ has a main role in evaluating the
temperature and its gradient inside the nuclear matter as seen by Eq.(4).
Fig.(3) shows the temperature as a function of $\overline{\eta }$ for the
reactions at 400 and 2100 MeV incident kinetic energy per nucleon. The
maximum temperature is found to be $55$ and $230$ $MeV$ respectively. The
proton density function in its equilibrium form is calculated according to
Eq.(6) over the effective range of the thermodynamic system. The results are
shown in Fig.(4) for protons emitted with $E_{L}=30,120$ and $180MeV$ with
emission Lab angle $30^{o}$. The protons produced at low energy show
anisotropic distribution with peaks near the origin and the surface of the
thermodynamic system. The position of the two peaks correspond to the
regions characterized by low $\overline{\eta }$ values and consequently low
temperature. High energy emission $(120-180MeV)$ shows plateau shape density
distribution. The bulk of which corresponds to high temperature zones. The
yield from the low temperature zones decreases with increasing the energy of
the emitted protons. The spatial variation of the function $f_{o}(\overline{r%
},\overline{p})$ is also studied. Fig.(5) exhibit the nth order derivative
of $f_{o}$. Maximum variation of the derivatives occurs at the origin and
near the surface where the temperature and its gradient also change rapidly.
The solution of the Vlasov equation is calculated to the fourth order
approximation as clarified by Eq.(8). The result is integrated over the
effective range with a geometrical weight factor $W(\eta )$, depends on the
size of the nuclear matter. Assuming azimuthal symmetry of the system then
one can finally find the Lab energy spectra, which is calculated at specific
values of emission angles $\theta =30,60,90,120$ and $150^{o}$.

\begin{equation}
f_{L}^{n}(E)=\int f^{n}(E,r)W(\eta (r))4\pi r^{2}dr
\end{equation}
Figs(6-11) shows the lab energy spectra of protons produced in Ne-U at 400 A
MeV corrected to the second order. The prediction of the model is compared
to the zero order correction (the equilibrium distribution) as well as the
experimental data. The emission time parameter is found by fitting method to
be $-12,-12,-8,-4.5$ and $-2(GeV)^{-1}$ corresponding to the emission angles 
$\theta =30,60,90,120$ and $150^{o}$ respectively. A global fair agreement
is obtained by the second order corrected solution of the Vlasov equation.
An appreciable improvement is observed in the prediction of the model,
particularly in the forward emission spectra ( $\theta $ = 30 and 60$^{o}$).
The prediction of the model comes closer to the equilibrium distribution for
the wide emission angles ( $\theta $ = 90, 120 and 150$^{o}$). Particle
produced at this wide angles are expected to make multiple collisions inside
the nuclear matter before emission takes place. In other words, the wide
angle production is a signal to the approach to the equilibrium state. Such
a system is characterized by large number of inter-nuclear cascade
collisions which increases the entropy of the system and leads to
equilibrium. It is found that the solution of the Vlasov equation through
the present approximation forms a converging series. It is enough to
consider only the first two terms in the series on dealing with the Ne-U
collisions at 400 A MeV. While terms up to the fourth order are found to
have appreciable values for the case of Ne-U collisions at 2100 A MeV. The
second correction and the fourth order correction improve the calculation
toward the virtual values. Although we considered four terms for the
reaction at 2100 MeV, but the agreement with experimental data was not fair
enough. The hypothesis of pre-equilibrium includes many approximations that
may not fit the systems formed at high energies. The last term in Vlasov
equation containing the field potential between the interacting particles
should be considered, and the problem then is treated microscopically
instead of the macroscopic picture as presented in this article. Despite of
the existence of the field theory of strong interactions, the theoretical
description of this phenomena is necessarily phenomenological because of the
very nature of the problem which involves many degrees of freedom. In a
forthcoming article, one may encounter the problem taking into consideration
the study of collective properties of hadronic matter, in particular its
possible phase transition to the quark-gluon phase. The main merits of this
approach lie, in our view, in the fact that the phenomenology is reduced to
microscopic concepts like parton- parton cross sections and structure
function and different equilibrium properties of the gluon and quark
components of hadrons which allow for identification of the coherent parts
of the interactions.

\section{ Conclusive remarks}

i- The pre-equilibrium model with reasonable approximations may fit the
experimental data of heavy ion collisions within the regime of few hundreds
A MeV. In this case, a power series is presented to describe the nuclear
density function in the frame of a thermodynamic picture.

ii- The relative projectile density plays an important role in determination
of the hadronic matter temperature.

iii- The temperature gradient in the system comes due to the assumption that
the nuclear matter has a diffuseness surface density with Gaussian
distribution.

iv- The temperature has minimum values around the center and near the end of
the effective range of the nuclear matter. These regions are responsible of
the emission of low energy particles, while fast particles are produced in
the bulk region characterized by $\eta =~0.5\quad $and high temperature.

v- The power series solution is converging in nature. The $n^{th}$ order
term depends on the $n^{th}$ derivative of the phase space distribution
function.

vi- A series up to the second order correction is sufficient to describe the
reaction at 400 A MeV. While the fourth order term has appreciable
importance in the reactions at 2100 A MeV.

vii- Particles emitted in the forward direction are produced in the early
stage of the reaction, far from the equilibrium. Backward production comes
in a later stage when the system constituents undergo multiple cascade
collisions.
\begin{verbatim}
 
 
\end{verbatim}

\begin{verbatim}
 
\end{verbatim}

\begin{center}
{\LARGE Figure Captions}
\end{center}

\begin{enumerate}
\item[Figure(1)]  Schematic diagram of the projectile and target collision
at a given impact

parameter $\stackrel{\_}{b}$.

\item[Figure(2)]  The relative projectile density $\eta $ for Ne-U collision
as measured from the center of the U target.

\item[Figure(3)]  The variation of the temperature as a function of the
relative projectile density $\eta $, for the Ne-U interactions at 400 and
2100 A MeV incident kinetic energies.

\item[Figure(4)]  The proton density spectra $f_{o}(r,p)$ at emission Lab
energies E$_{L}$= 30, 120,180 MeV and Lab angle $\theta _{L}$= 30$^{o}$
produced in Ne-U collision at 400 MeV kinetic energy.

\item[Figure(5)]  The n$^{th}$ order derivatives of the proton density
spectra $f_{o}(r,p)$. The first three derivative at emission Lab energy E$%
_{L}$= 30 MeV and Lab angle $\theta _{L}$= 30, produced in Ne-U collision at
400 MeV.

\item[Figure(6)]  The energy spectra of protons produced in Ne-U
interactions at 400 A MeV, at

emission angle 30$^{o}$. The solid line represents the equilibrium
distribution, while the

second correction is represented by dashed line, with an emission time
parameter $t=-12$ $(GeV)^{-1.}$

\item[Figure(7)]  The energy spectra of protons produced in Ne-U
interactions at 400 A MeV, at

emission angle 60$^{o}$. The solid line represents the equilibrium
distribution, while the

second correction is represented by dashed with an emission time parameter $%
t=-12$ $(GeV)^{-1.}$

\item[Figure(8)]  The energy spectra of protons produced in Ne-U
interactions at 400 A MeV, at

emission angle 90$^{o}$. The solid line represents the equilibrium
distribution, while the second correction is represented by a dashed line,
with an emission time parameter $t=-8$ $(GeV)^{-1.}$

\item[Figure(9)]  The energy spectra of protons produced in Ne-U
interactions at 400 A MeV, at

emission angle 120$^{o}$. The solid line represents the equilibrium
distribution, while the second correction is represented by dashed line with
an emission time parameter $t=-4.5$ $(GeV)^{-1.}$

\item[Figure(10)]  The energy spectra of protons produced in Ne-U
interactions at 400 A MeV, at

emission angle 150$^{o}$. The solid line represents the equilibrium
distribution, while the second correction is represented by dashed line with
an emission time parameter $t=-2.5$ $(GeV)^{-1.}$

\item[Figure(11)]  The second order corrected pre-equilibrium energy spectra
of protons produced

in Ne-U interactions at 400 A MeV, at emission angles of 30, 60, 90, 120 and
150$^{o}$. The solid lines represent the calculated distributions and the
experimental data are represented by (+) signs.
\end{enumerate}

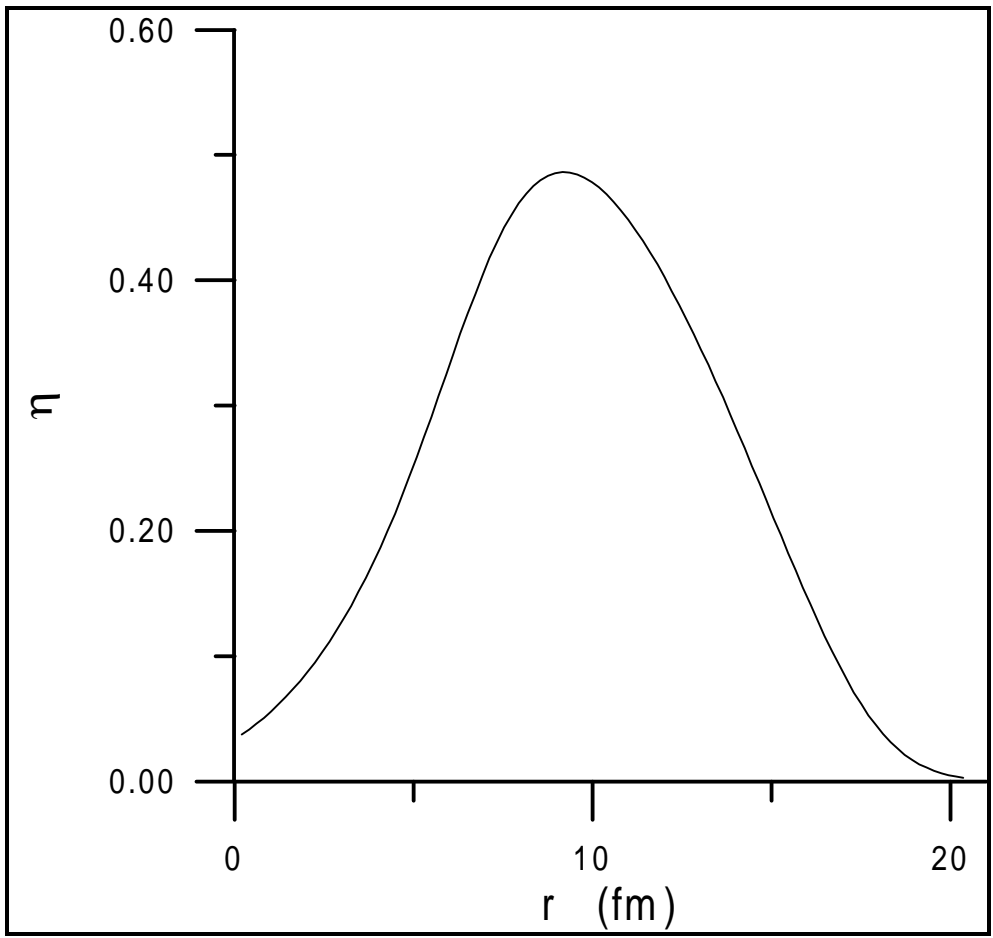
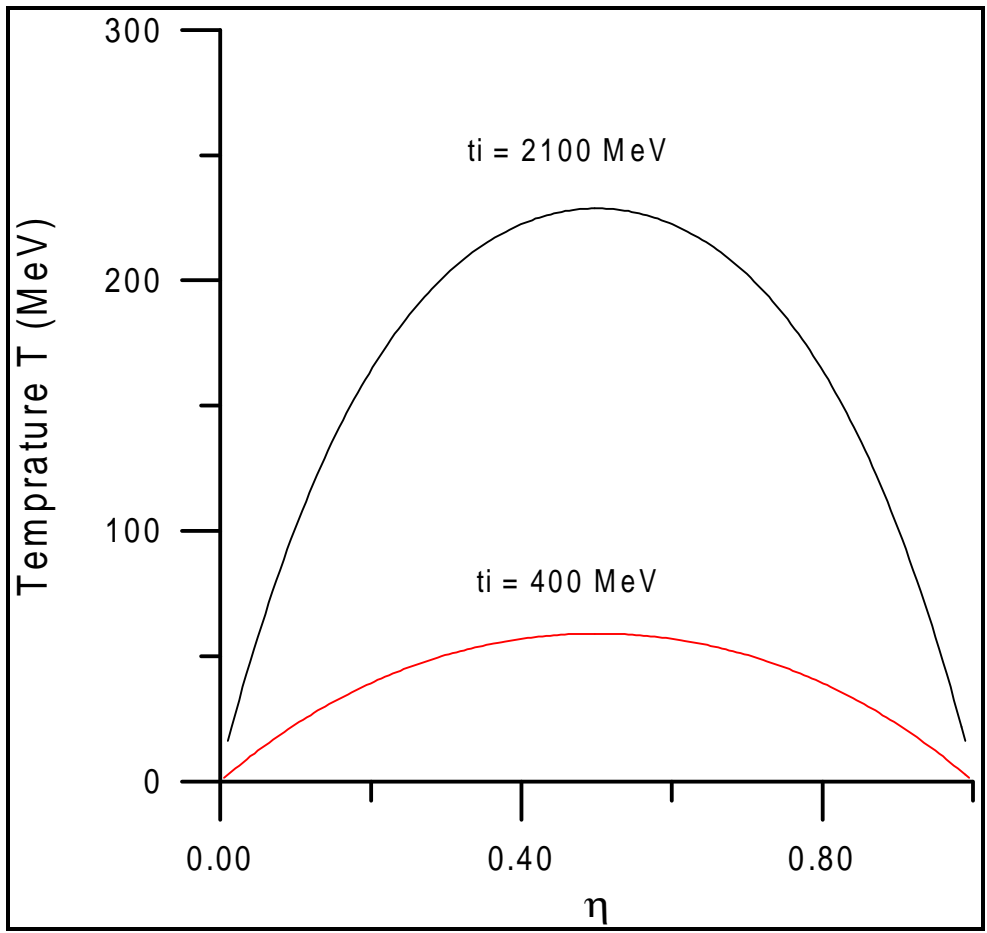
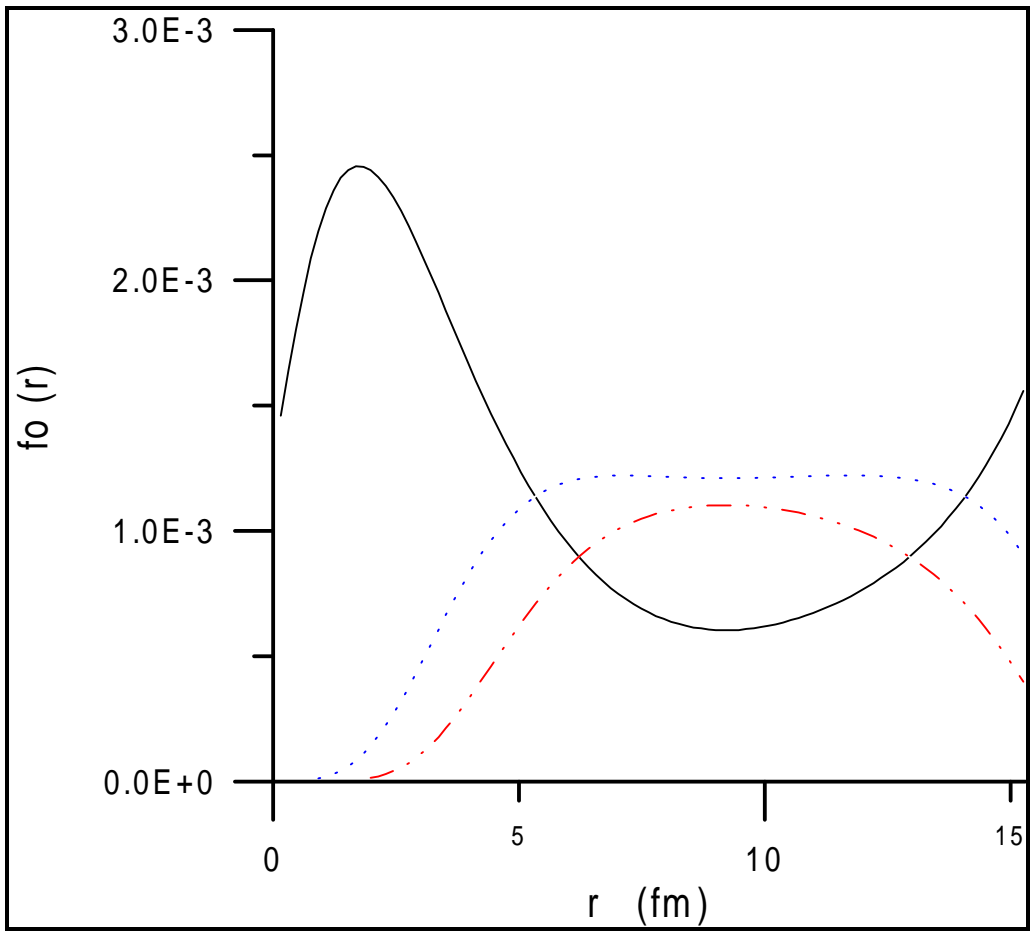
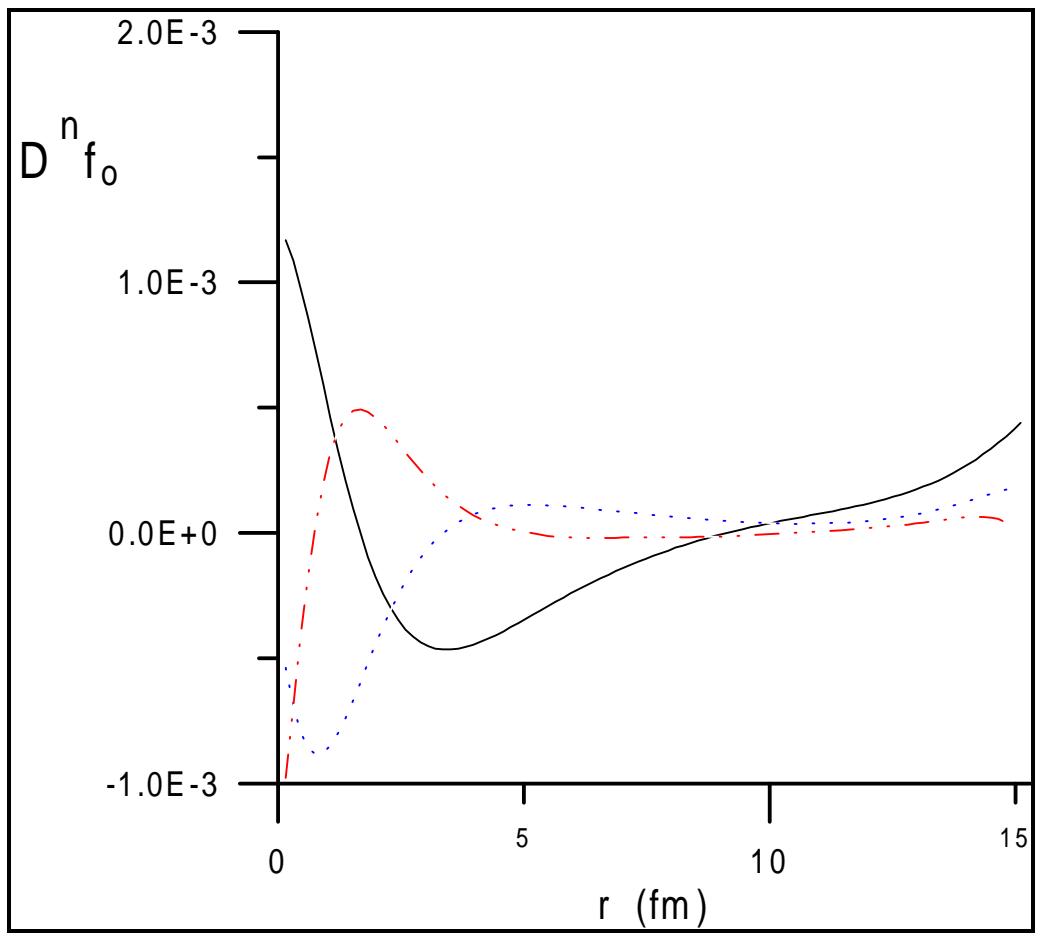
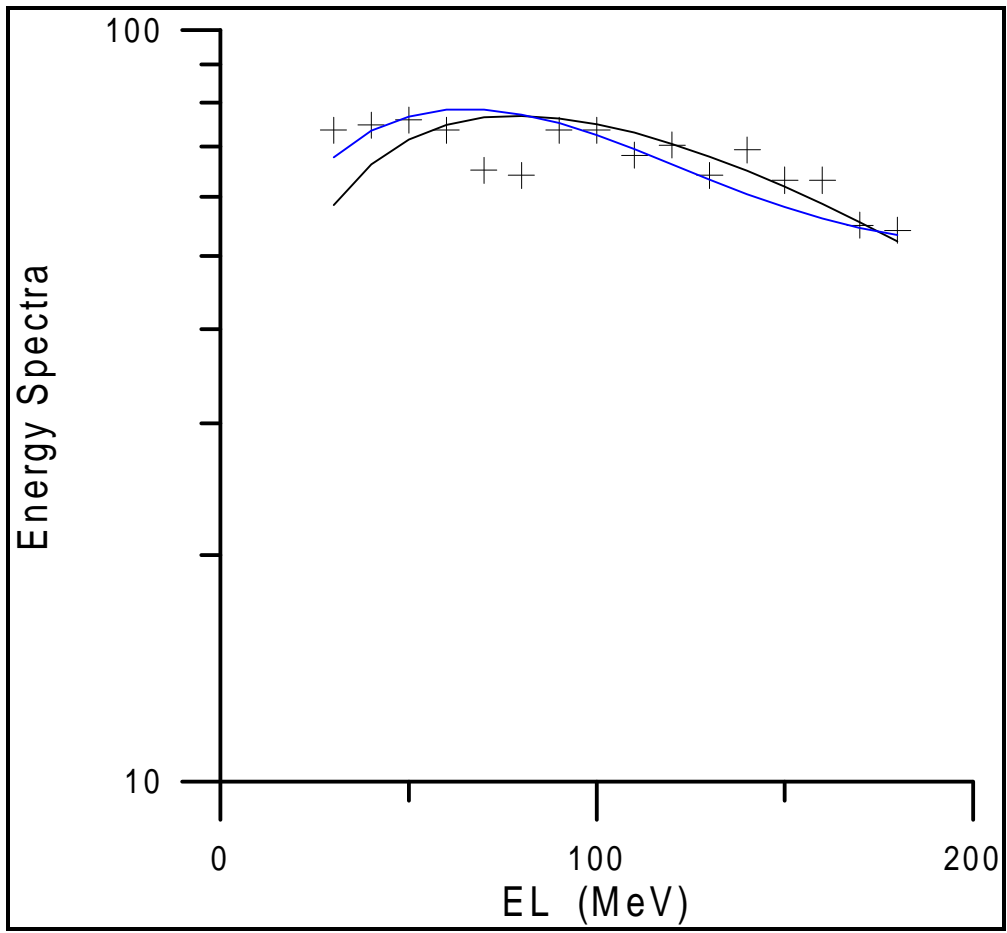
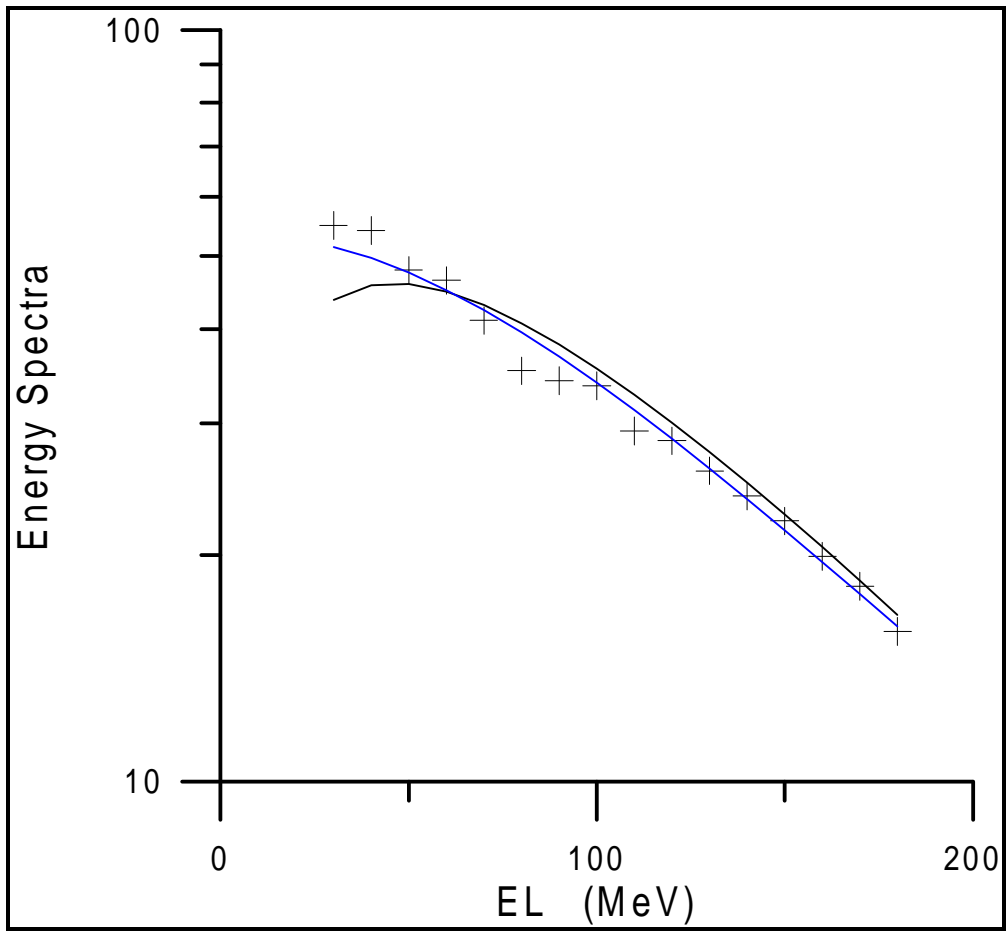
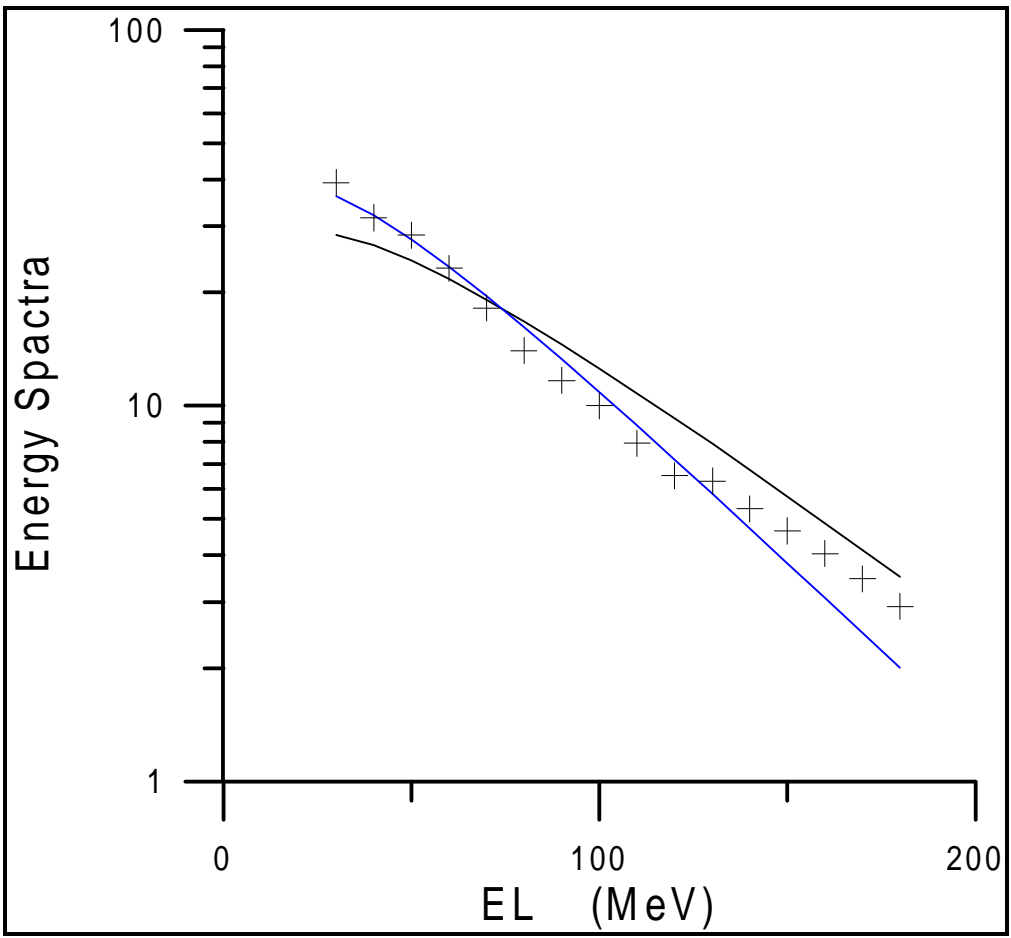
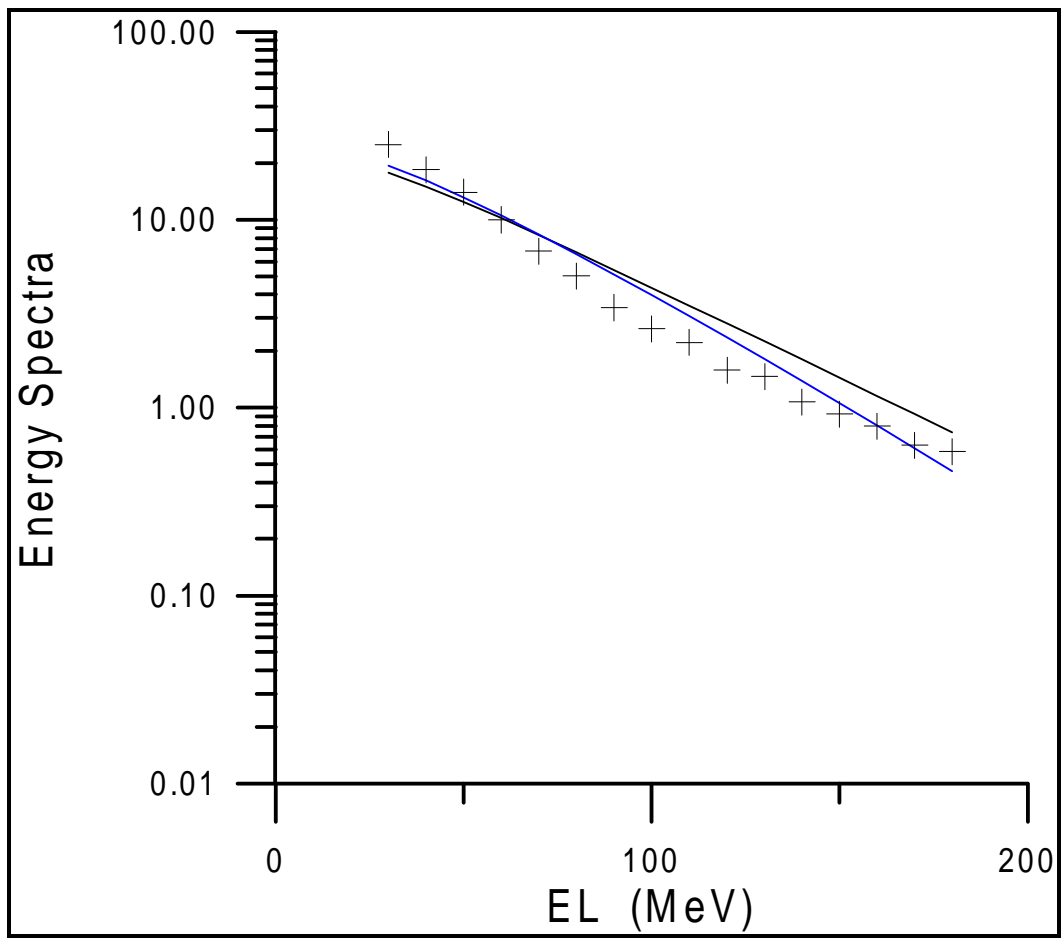
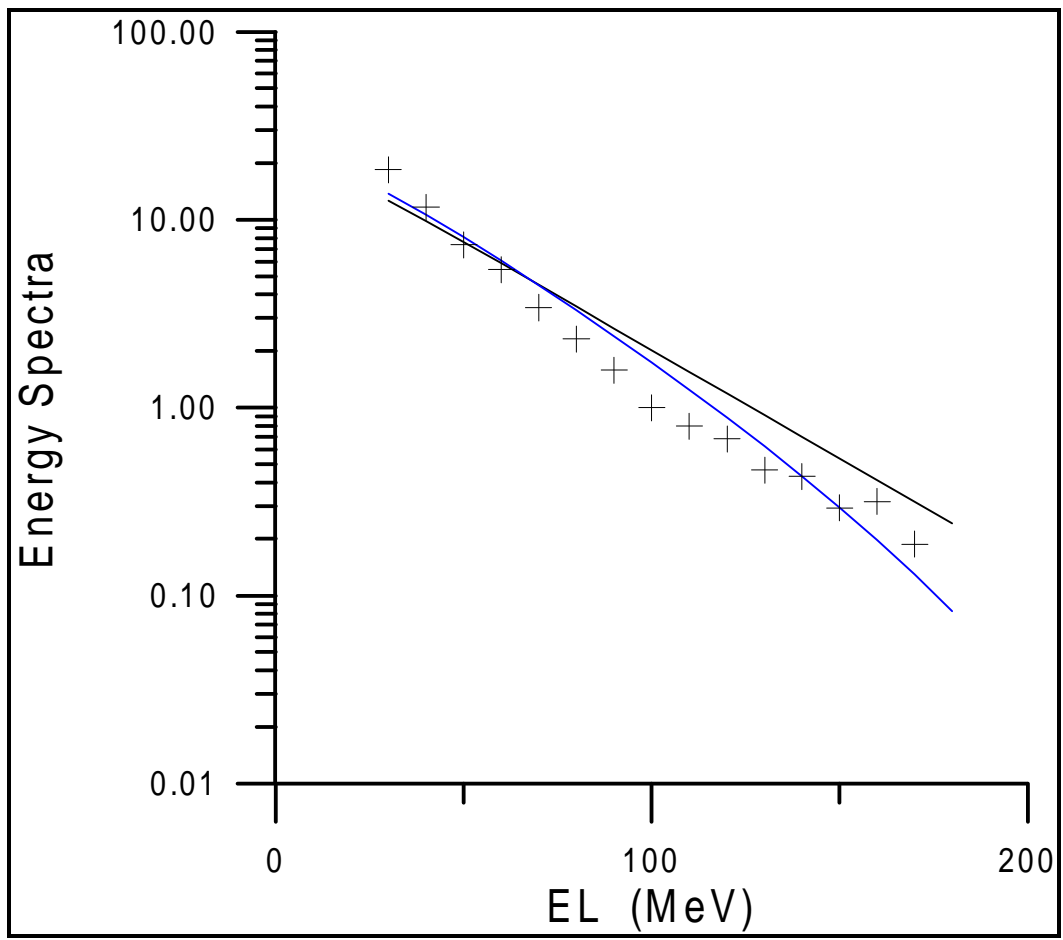
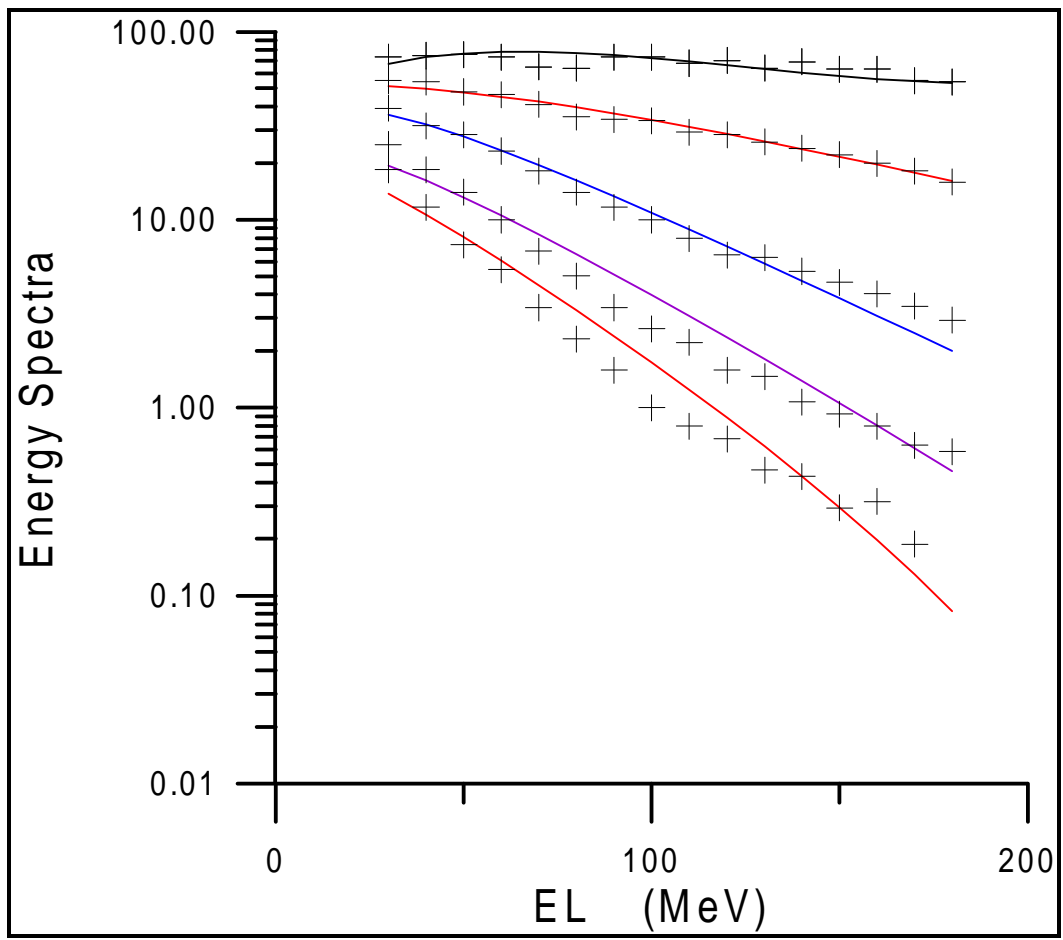
\end{document}